# "Suspended" Pt nanoparticles over TiO$_2$ nanotubes for enhanced photocatalytic H$_2$ evolution

Nhat Truong Nguyen,[a] JeongEun Yoo,[a] Marco Altomare,[a] Patrik Schmuki[a,b]*

[a] Department of Materials Science and Engineering WW4-LKO, University of Erlangen-Nuremberg, Martensstrasse 7, D-91058 Erlangen, Germany.
[b] Department of Chemistry, King Abdulaziz University, Jeddah, Saudi Arabia
* Corresponding author. E-mail: schmuki@ww.uni-erlangen.de
Fax: +49-9131-852-7582; Tel: +49-9131-852-7575

# A B S T R A C T

In the present work we introduce a technique to form a photocatalyst, based on Pt nanoparticles suspended over the mouth of TiO$_2$ nanotubes. These structures are obtained by the decoration of the top end of highly ordered TiO$_2$ nanotubes with a web of TiO$_2$ nanofibrils, followed by sputter deposition of a minimum amount of Pt. A subsequent thermal dewetting step forms 3–6 nm-sized Pt nanoparticles along the nanofibrils. These structures, when compared to conventional Pt decoration techniques of TiO$_2$ nanotubes, show a strongly enhanced photocatalytic H$_2$ evolution efficiency.

Ever since the ground-breaking report by Fujishima and Honda in 1972, $TiO_2$ has been increasingly investigated for the conversion of solar light into electrical and chemical energy (*e.g.*, photocatalysis, photo-electrochemistry).[1-3] A key advantage of $TiO_2$ over other semiconductive systems is that except for the proper matching of band-edge positions with respect to $H_2O$ red-ox potentials, it has a very high chemical stability that almost completely suppresses photocorrosion. In aqueous environments, under UV or solar light irradiation, electrons and holes generated in the $TiO_2$ conduction and valence bands, respectively, are able to split water into $H_2$ and $O_2$.[4-6]

Usually $TiO_2$ photocatalysts are mainly based on powders. However, one-dimensional morphologies, such as nanowires, nanorods and nanofibers were recently shown to exhibit preferential percolation pathways for charge carrier separation, hence resulting in highly photoactive materials.[7]

One of the most investigated structures over the past few years are ordered $TiO_2$ nanotube arrays formed by anodization of Ti substrates in adequate electrolytes. Already early works of Assefpour-Dezfuly *et al.* and Zwilling *et al.* provided the basis for the growth of $TiO_2$ nanotubes by anodizing Ti in dilute fluoride-containing electrolytes.[8,9] Afterwards, growth conditions have been widely investigated in order to fabricate $TiO_2$ nanotubes with a large range of different structural and morphological features.[3] Anodic $TiO_2$ nanostructures not only have shown favorable electron transport but also have been reported to show a better photocatalytic activity compared to $TiO_2$ nanoparticle layers.[10-12]

Nevertheless, if any $TiO_2$ structures are used towards photocatalysis for water splitting, the deposition of a co-catalyst is required to achieve a reasonable $H_2$ efficiency. In particular, the deposition of Pt nanoparticles on $TiO_2$ leads to a drastically more effective $H_2$ production. This is typically ascribed to following two reasons: *i)* Pt nanoparticles are able to trap conduction band electrons and to mediate their transfer to the liquid phase; *ii)* Pt sites represent efficient catalytic sites for the recombination of atomic hydrogen to $H_2$.[4,13-17] $TiO_2$ structures are thus decorated with Pt using photo-deposition, electro-deposition, sol-gel and impregnation.[18-20]

In the present work, we introduce a novel Pt@$TiO_2$ nanotube platform that provides higher $H_2$ evolution efficiencies than conventional approaches. Our goal was an efficient decoration of $TiO_2$ with a minimum amount of Pt. For this we present an approach that leads to nanoparticles suspended over the top opening of $TiO_2$ nanotubes as shown in Fig. 1. To fabricate this structure we used a processing sequence as outlined in scheme 1. Tubes, grown by electrochemical anodization, were firstly subjected to a simple chemical treatment in alkaline solution which allowed for the formation of a $TiO_2$ nanofibril web at the mouth of the tubes. Then, sputter-deposition was employed to decorate the nanofibrils with minimal amounts of Pt. The Pt decoration then was converted into 3-6 nm-sized Pt nanoparticles by a thermal dewetting step. The photocatalytic $H_2$ evolution efficiency from these structures then was compared to conventional Pt decoration techniques of $TiO_2$ nanotubes.

Fig. 1 shows SEM and TEM images of the nanostructures at each stage of their fabrication. Highly ordered $TiO_2$ nanotubes (see Fig. S1) were produced by anodization of Ti in hot $HF/H_3PO_4$ mixtures (Fig. 1(a) and (b)). These nanotubes have a diameter of ~ 80 nm and a height of ~ 200 nm, and show a reaction-vessel geometry that was demonstrated to be optimal for UV light-driven photocatalysis.[21,22]

If these tube layers (after a first mild annealing in air) are exposed to a strongly alkaline solution (4 M $NaOH_{aq}$), $TiO_2$ nanofibrils form across the openings of the tubes (Fig. 1(c)).[23,24] In our case, this results in a web-like structure that consists of *ca.* 5 nm-wide fibrils (Fig. 1(h)). It should be pointed out that the air-annealing step is of crucial importance. As formed amorphous nanotubes were strongly damaged after a 30 min-NaOH soaking (with only a negligible nanofibril formation) and totally destroyed after 24 h (Fig. S2 (c)-(e)).

Fig. 1(d)-(h) illustrate the top and cross-section views of the structures produced after depositing a 1 nm-thick Pt layer followed by thermal dewetting. Only very little difference can be observed when comparing the SEM pictures of the structures before and after Pt deposition (Fig. 1(c)-(e)), this suggests that sputtering forms a rather conformal polycrystalline Pt layer which homogeneously coats the nanofibrils. On the other hand, the formation of Pt nanoparticles on the nanofibrils becomes apparent after a subsequent annealing in Ar atmosphere (450 °C – 30 min) (Fig. 1(f) and (g)). The agglomeration of the Pt to nanoparticles is due to thermal dewetting. The Pt particles are a few nm in diameter (3-6 nm) and their size is related to the nominal thickness of the initially deposited Pt layer. In particular, the mean diameter of the Pt nanoparticles increases by increasing the thickness of the initial Pt layer (Fig. S3 and S4). When the nominal Pt thickness reaches a value of *ca.* 10-15 nm, the Pt nanoparticles tend to agglomerate forming a Pt layer that coats the nanotube walls and leaves nearly no Pt deposits on the fibrils (Fig. S3(f)-(g)). Moreover, the cross-sectional TEM image in Fig. 1(i) shows that the Pt nanoparticles are preferentially attached to the nanofibrils and suspended over the tube opening (the HR-TEM image in Fig 1(j) shows a lattice constant of 2.22 Å corresponding well to a Pt (111) crystallographic plane).

Fig. 2(a) shows XRD patterns that were collected for the structures at different stages of their fabrication. The as-formed tubes are amorphous while the first heat treatment in air induces the crystallization of $TiO_2$ into a mixed anatase-rutile phase. The XRD patterns recorded for the Pt-decorated structures do not show any Pt peak when the nominal thickness of the sputtered Pt film is lower than 10 nm (such amounts of Pt might be undetectable by XRD). However, when the Pt film is 15 nm thick, Pt peaks (peaking at $2\Theta = 45.9°$ and $67.1°$) clearly appear in the diffractograms. When 15 nm thick Pt decorated $TiO_2$ nanotubes undergo the second thermal treatment (in Ar, at 450 °C for 30 min), the intensity of Pt peaks clearly increases. This might be related to a higher degree of crystallinity of the Pt nanoparticles for the double-annealed materials. It has been also reported that Pt sputtered layers might show a relevant content of amorphous Pt oxides, *i.e.*, Pt(II) and Pt(IV) oxides.[25] Therefore, the second annealing in Ar atmosphere may induce the reduction of Pt oxides into $Pt^0$ along with the crystallite growth.

XPS characterization (Fig. 2(b)) was performed for the different structures, before and after Pt deposition, to examine the chemical state of the samples. The actual deposition of Pt is confirmed by the appearance of signals peaking at 72.06, 314.86 and 331.66 eV, these values corresponding to the binding energies of Pt 4f, 4d5 and 4d3, respectively. A signal peaking at 493.1 eV indicates the presence of Na after the sample treatment in NaOH.

Fig. 3(a) shows the amount of $H_2$ produced from water-ethanol solutions under UV illumination (laser, 325 nm, 60 mW cm$^{-2}$) for structures decorated with different Pt amounts (nominal thickness in the 0-15 nm range).[4,15,18,19,26] While the bare structures (*i.e.*, no Pt decoration) only lead to a production of 2.8 µL $H_2$ / 9 h, the Pt-decorated structures show, as expected, a much higher photocatalytic $H_2$ production rate ($r_{H2}$). We observe that the $r_{H2}$ drastically increased when a few nm-thick Pt films are deposited. Pt thickness of 1 nm leads to the highest $r_{H2}$ of 873 µL (*i.e.*, *ca.* 0.1 mL h$^{-1}$). In other words, only minimal amounts of Pt are needed in the suspended geometry to efficiently place the catalyst. For this catalyst, the amount of evolved $H_2$ linearly increases with irradiation time, *i.e.*, the $r_{H2}$ is steady over time, confirming that neither release (fall-off) of Pt nanoparticles nor (photo-)corrosion of the structures takes place (Fig. S5). The structures decorated with nominal a layer of 1 nm Pt already represent an optimum in view of photocatalytic $H_2$ production. When the Pt amounts are increased, a considerable drop of photocatalytic activity occurs. This may be due to *i)* a large increase of the mean size of Pt nanoparticles and *ii)* "shading effect", *i.e.*, larger amounts of deposited Pt optically shield the underneath structure so that the semiconductor is actually exposed to a lower specific photon flux (Fig. S3).

Fig. 3(b) shows a comparison of the photo-activity of different $TiO_2$ structures decorated with a 1 nm-thick Pt film. Both Pt@compact $TiO_2$ and Pt@$TiO_2$ nanotubes show a much lower $r_{H2}$ compared to nanotubes decorated with suspended Pt nanoparticles (Fig. S6). The low efficiency of the compact film is clearly due to its low surface area. On the other hand, the low $r_{H2}$ of Pt@$TiO_2$ nanotubes demonstrates that the nanofibrils represent a significant geometry improvement that drastically improves the photocatalyst performance for the same Pt loading. This may be explained by short charge carrier diffusion length in the fibrils. It is noteworthy that also the NaOH-soaking time affects $r_{H2}$: a 30 min-long treatment leads to the most efficient photocatalyst while a loss of activity was observed when fibrils are too packed (60 min-long soaking), probably because of hindered light absorption (*i.e.*, the inner fibrils might be shadowed by the outermost ones) or because of larger Pt nanoparticle formation (Fig. S7).

Another crucial point affecting the efficiency of the structures is the annealing treatment. As shown in Fig. 3(b), the tubes decorated with suspended Pt nanoparticles showed a doubling of $r_{H2}$ after a second Ar-annealing. The reason for this is that as-formed fibrils are amorphous, and annealing is required to achieve conversion into crystalline oxide. Besides crystallization, the second Ar-annealing also causes the Pt dewetting and formation of Pt nanoparticles. The Ar-atmosphere during the second annealing is crucial. Experiments in where the second heat treatment was conducted in air (Fig. 3(b)) showed only a negligible increase of $r_{H2}$ compared to single-annealed structures. Additional experiments were carried out using different treatment times of the second Ar-annealing or a different sputtering-annealing sequence (Fig. S8). In the first case, the highest $r_{H2}$ was obtained with a 30 min-long Ar-treatment, this probably due to *i)* an optimized mean size and crystallinity of the Pt nanoparticles, and *ii)* optimized light absorption ability of the fibrils. If the second Ar-annealing is performed before Pt sputtering, a markedly low $r_{H2}$ is obtained. Therefore, one might conclude that Ar-annealing allows dewetting and induces crystallite growth. Pt-dewetting partially exposes the fibrils, *i.e.*, it uncovers them and a larger photon flux can be absorbed.

Overall, in the present work we introduce the fabrication of a very efficient photocatalyst geometry that is based on Pt nanoparticles suspended over the top opening of highly ordered TiO2 nanotubes. Their preparation is based on simple self-ordering processes. Only minimum amount of Pt is required to achieve strongly enhanced photocatalytic H2 evolution efficiencies compared with conventional Pt decoration approaches.

The authors would like to acknowledge ERC, DFG and the DFG cluster of excellence EAM for financial support as well as H. Hildebrand for valuable technical help.

**Figure captions**

**Scheme. 1** Formation of suspended Pt nanoparticles over $TiO_2$ nanotubes: (i) anodization in hot $H_3PO_4$/HF electrolyte, (ii) first annealing in air and NaOH soaking, (iii) Pt sputtering and (iv) second annealing in Ar.

**Fig. 1** SEM images of a), b) as-formed $TiO_2$ NTs; c) nanofibrils after 30 min-long NaOH-treatment; 1 nm-thick suspended Pt nanoparticles: d), e) before and f), g), h) after thermal dewetting in Ar; i) TEM image of suspended Pt nanoparticles and j) HRTEM image of Pt nanoparticles showing a lattice constant of 2.22 Å corresponding to the (111) crystallographic plane.

**Fig. 2** a) XRD patterns of (i) as-formed $TiO_2$ nanotubes; (ii) annealed $TiO_2$ nanotubes; (iii) 1 nm-thick Pt suspended over $TiO_2$ nanotubes after thermal dewetting in Ar; 15 nm-thick Pt suspended over $TiO_2$ nanotubes (iv) before and (v) after thermal dewetting in Ar. b) XPS spectra of $TiO_2$ nanotubes with nanofibrils and suspended Pt nanoparticles (NPs) over $TiO_2$ nanotubes (NTs). The inset shows the enlarged area of the XPS spectra of Pt4f.

**Fig. 3** Photocatalytic $H_2$ evolution measured for: a) $TiO_2$ nanotubes decorated with different nominal thicknesses of Pt; b) different types of $TiO_2$ structures all decorated with 1 nm-thick Pt layer. All experiments lasted 9 h and were carried out under UV light irradiation (HeCd laser, $\lambda = 325$ nm, power = 60 mW cm$^{-2}$).

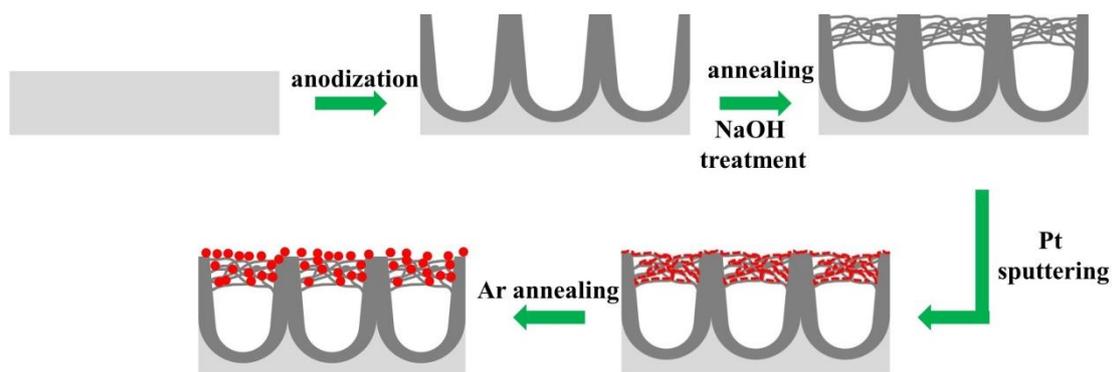

**Scheme 1.** Formation of the suspended Pt nanoparticles over TiO$_2$ nanotubes: (i) anodization in hot H$_3$PO$_4$/HF electrolyte, (ii) first annealing in air and NaOH soaking, (iii) Pt sputtering and (iv) second annealing in Ar.

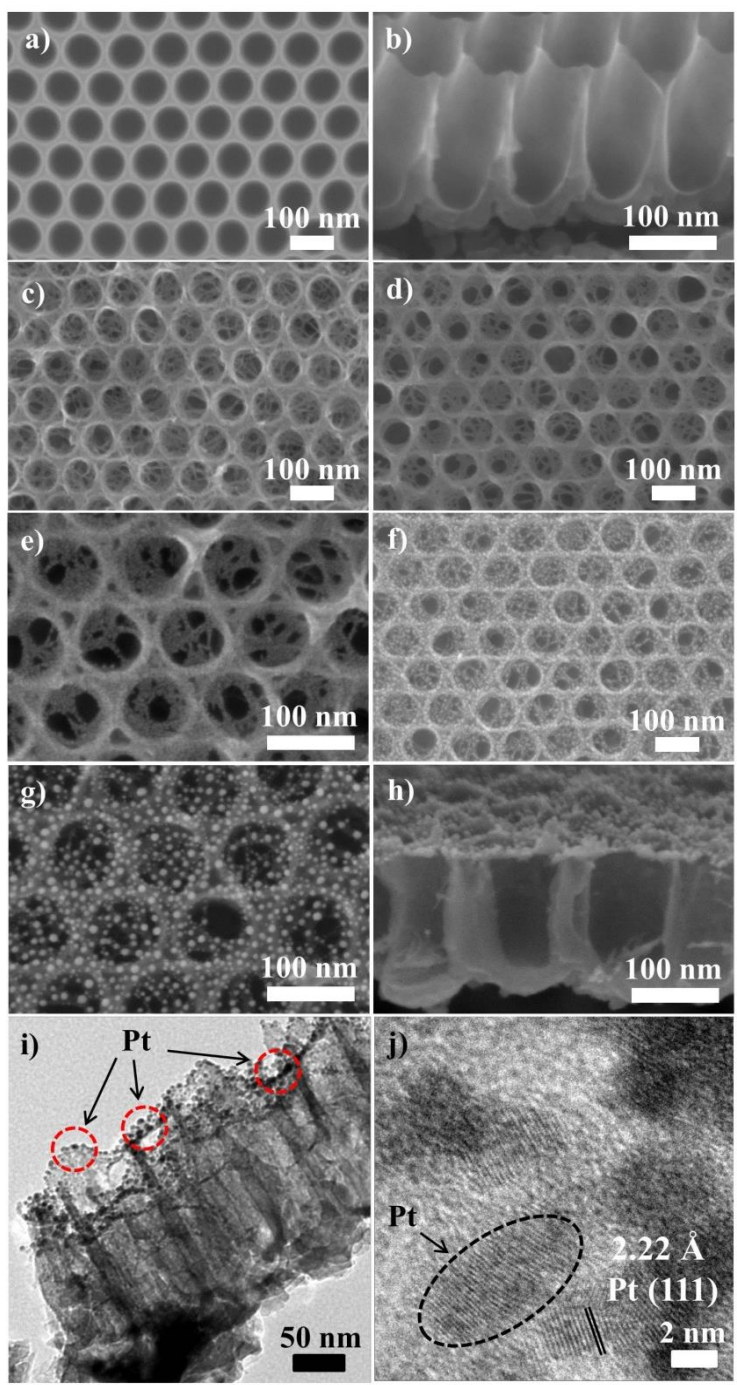

**Figure 1.** SEM images of a), b) as-formed TiO$_2$ nanotubes; c) nanofibrils after 30 min-long NaOH-treatment; 1 nm-thick suspended Pt nanoparticles: d), e) before and f), g), h) after thermal dewetting in Ar; i) TEM image of suspended Pt nanoparticles and j) HRTEM image of Pt nanoparticles showing a lattice constant of 2.22 Å corresponding to the (111) crystallographic plane.

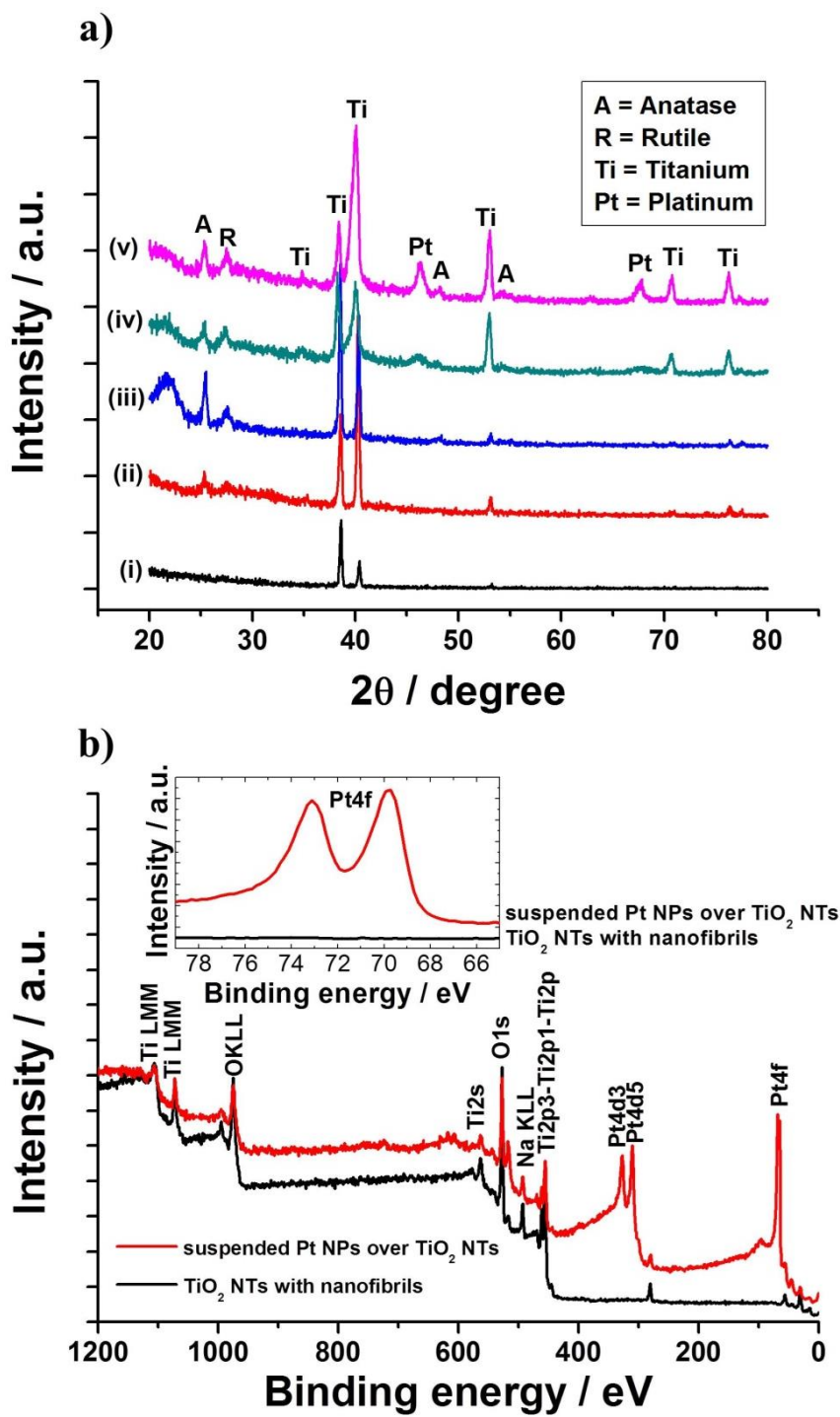

**Figure 2.** a) XRD patterns of (i) as-formed TiO$_2$ nanotubes; (ii) annealed TiO$_2$ nanotubes; (iii) 1 nm-thick Pt suspended over TiO$_2$ nanotubes after thermal dewetting in Ar; 15 nm-thick Pt suspended over TiO$_2$ nanotubes (iv) before and (v) after thermal dewetting in Ar. b) XPS spectra of TiO$_2$ nanotubes with nanofibrils and suspended Pt nanoparticles (NPs) over TiO$_2$ nanotubes (NTs). The inset shows the enlarged area of the XPS spectra of Pt4f.

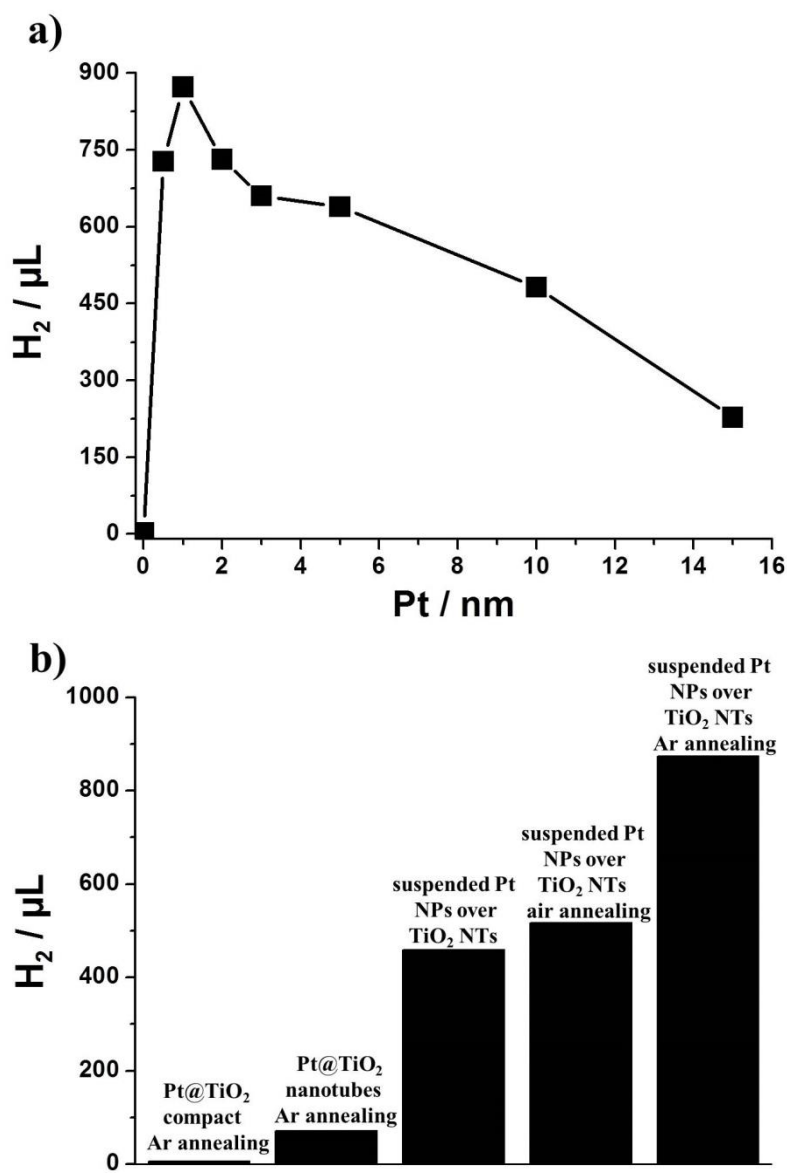

**Figure 3.** Photocatalytic H$_2$ evolution measured for: a) TiO$_2$ nanotubes decorated with different nominal thicknesses of Pt; b) different types of TiO$_2$ structures all decorated with 1 nm-thick Pt layer. All experiments lasted 9 h and were carried out under UV light irradiation (HeCd laser, $\lambda$ = 325 nm, power = 60 mW cm$^{-2}$).

**Electronic Supplementary Information for**

# "Suspended" Pt nanoparticles over TiO$_2$ nanotubes for enhanced photocatalytic H$_2$ evolution

Nhat Truong Nguyen, JeongEun Yoo, Marco Altomare, Patrik Schmuki*

Department of Materials Science and Engineering WW4-LKO, University of Erlangen-Nuremberg, Martensstrasse 7, D-91058 Erlangen, Germany.
*Corresponding author. Email: schmuki@ww.uni-erlangen.de

## Table of contents:

**Growth of TiO$_2$ nanotubes**

Titanium foils (0.125 mm thickness, 99.6+ % purity, Advent Research Materials) were decreased in acetone, ethanol and deionized water and then dried in N$_2$ stream. Anodization was conducted in a simple two-electrode electrochemical cell, in a hot H$_3$PO$_4$/HF electrolyte, at 15 V for 2 h, using a DC power supply (VLP 2403 pro, Voltcraft). The concentration of HF was 3 M and all the anodization experiments were performed by fixing the temperature of the electrolyte typically at 80 °C (these experimental conditions are similar to those reported in some of our previous works).[1-3] Ti pieces and a platinum foil were used as working and counter electrodes, respectively. After anodization, the foils were rinsed with ethanol and subsequently dried in a N$_2$ stream. Subsequently, the samples were annealed at 450 °C for 30 min in air using a Rapid Thermal Annealer (Jipelec Jetfirst 100 RTA), with a heating and cooling rate of 30 °C min$^{-1}$.

**Fabrication of TiO$_2$ nanofibrils**

The nanofibrils were formed at the opening of the tubes simply by soaking the TiO$_2$ nanotube arrays in a 4 M NaOH aqueous solution for 30 min at room temperature, in the absence of stirring. Then, the samples were taken off, washed with deionized water, and dried in N$_2$ stream.

Test experiments carried out by using another alkaline electrolyte (KOH$_{aq}$ instead of NaOH$_{aq}$) led to similar structure of the nanofibrils (Fig. S2(a)), therefore proving that the formation of the fibrils is simply ascribed to a localized TiO$_2$ dissolution-precipitation reaction induced by the high concentration of hydroxide anions. We also observed that the formation of the fibrils is suppressed if the NaOH treatment is performed under stirring (Fig. S2(b)). Nevertheless, it is important to notice that the nanofibrils formation occurs without affecting the structure and the ordered alignment of the nanotubes.

We also performed some NaOH treatments on TiO$_2$ nanotubes fabricated by conventional electrochemical anodization, *i.e.*, grown in glycerol- or ethylene glycol-based electrolytes. Noteworthy, the formation of nanofibrils did not take place when employing these nanotubes, this most likely owing to a different chemical composition of the tubes grown in organic-based electrolytes compared to ones grown in hot HF/H$_3$PO$_4$ solutions.

**Pt decoration**

In order to decorate Pt on the TiO$_2$ nanotubes, plasma-sputtering (EM SCD500, Leica) was used to deposit Pt thin films. Different Pt amounts were sputtered on the nanofibrils-into-tubes substrates by choosing various nominal thicknesses in the range between 0.5 – 15 nm. The deposition was usually carried out at 10$^{-2}$ mbar of Ar, by applying a current of 17 mA and by using a 99.99 % pure Pt target (Hauner Metallische Werkstoffe). The average coating rate was 0.07 nm s$^{-1}$. After Pt decoration, the samples were annealed (and dewetted) in Ar, at 450 °C for 30 min.

## Characterization of the structures

For morphological characterization of the samples, a field-emission scanning electron microscope (FE-SEM, Hitachi S4800) and high resolution transmission electron microscope (HR-TEM, Philips CM300) were employed. The crystallographic properties of the materials were analyzed by X-ray diffraction (XRD) performed with a X′pert Philips MPD (equipped with a Panalytical X'celerator detector) using graphite monochromized Cu Kα radiation ($\lambda$ = 1.54056 Å). X-ray photoelectron spectroscopy (XPS, PHI 5600, US) was used to characterize the chemical composition of the samples.

**Photocatalytic experiments**

The photocatalytic $H_2$ production experiments were conducted by immersing the samples in a 20 vol.% ethanol aqueous solution and by irradiating them with UV light from a HeCd laser, Kimmon, Japan ($\lambda$ = 325 nm; beam size = 0.785 cm$^2$; irradiation power = 60 mW cm$^{-2}$). The experiments were run in a quartz tube photocatalytic cell, sealed up with a rubber septum from which 200 µL samples were withdrawn and analyzed to determine the amount of evolved $H_2$ by gas chromatography (GCMS-QO2010SE, Shimadzu). The GC was equipped with a thermal conductivity detector (TCD), a Restek micropacked Shin Carbon ST column (2 m x 0.53 mm) and a Zebron capillary column ZB05 MS (30 m x 0.25 mm). GC measurements were carried out at a temperature of the oven of 45 °C (isothermal conditions), with the temperature of the injector set up at 280 °C and that of the TCD fixed at 260 °C. The flow rate of the carrier gas, *i.e.*, argon, was 14.3 mL min$^{-1}$. All the experiments lasted 9 hours and the evolved $H_2$ was measured always at the end of the experiments (*i.e.*, after 9 h) aside from few checking runs during which gas samples were withdrawn after 1, 3 and 6 h to verify that the $H_2$ evolution was steady over time.

Photocatalytic experiments were carried out in ethanol-water mixtures since it has been proved that the presence of specific amounts of organics (*e.g.*, methanol, ethanol and glycerol) markedly triggers the $H_2$ production.[1,2,4-6] Precisely, ethanol acts as a hole-scavenger, meaning that the organic molecules are quickly oxidized toward several intermediate compounds and finally to $CO_2$. As a consequence of the fast hole-consumption, conduction band electrons are more readily available for water reduction, thus yielding overall larger amount of $H_2$.

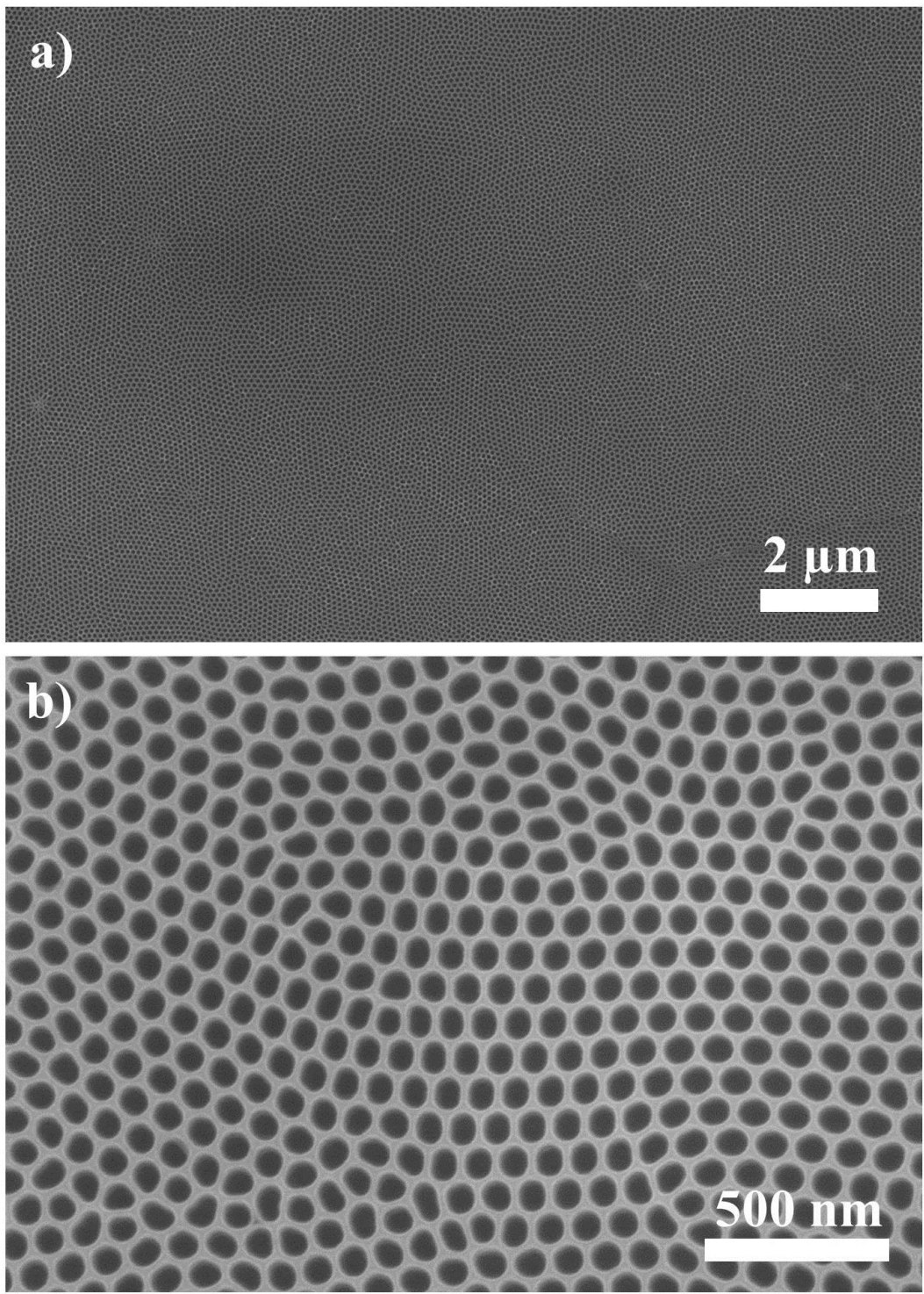

**Figure S1.** Top view SEM pictures of the highly ordered TiO$_2$ nanotubes taken at magnification of a) 8 K and b) 50 K.

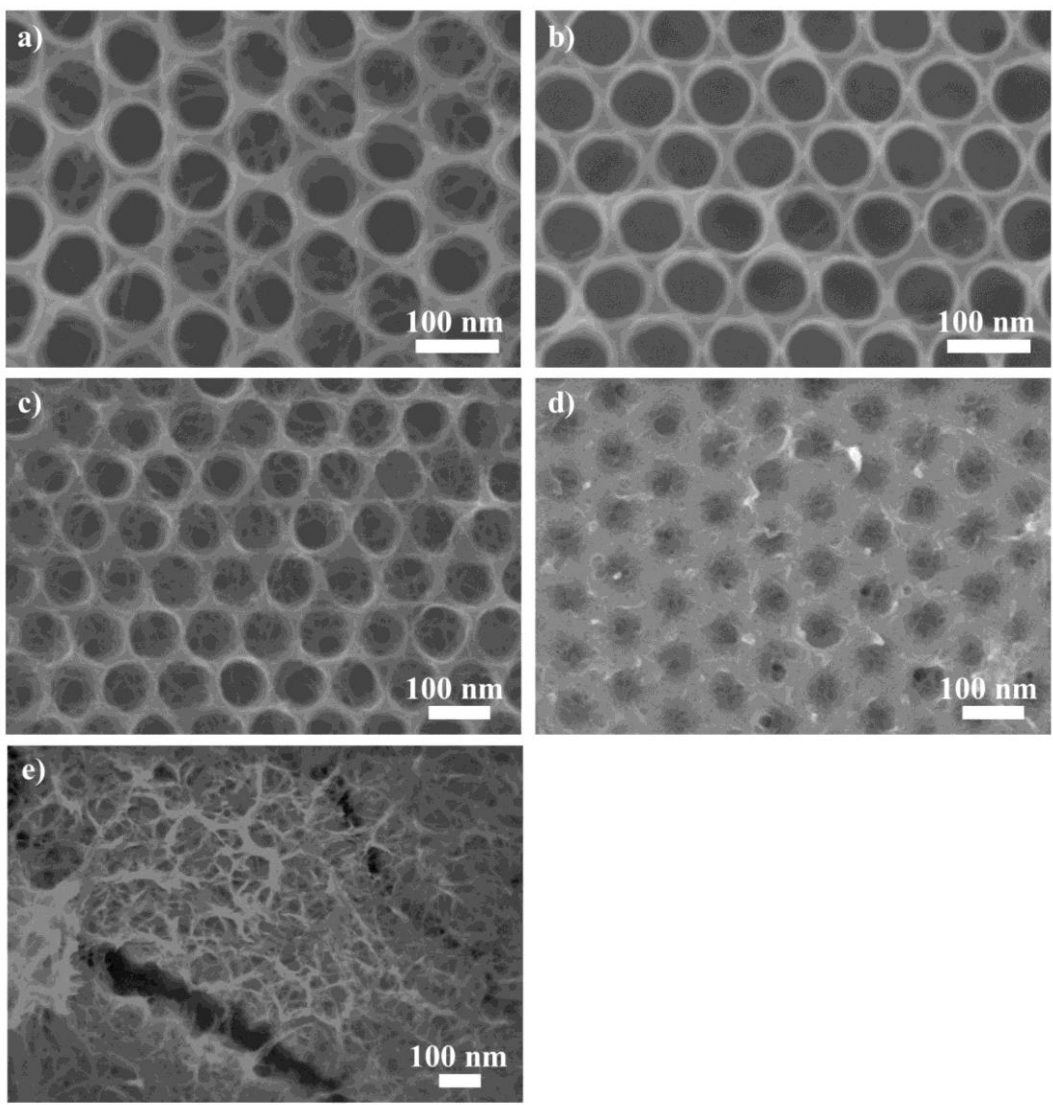

**Figure S2.** SEM images of nanofibrils formed: a) after a 30 min-long KOH-treatment; b) after a 30 min-long NaOH-treatment performed under stirring; c) after a 24 h-long NaOH-treatment; nanofibrils formed by performing d) 30 min- and e) 24 hours-long NaOH-treatment on amorphous nanotubes.

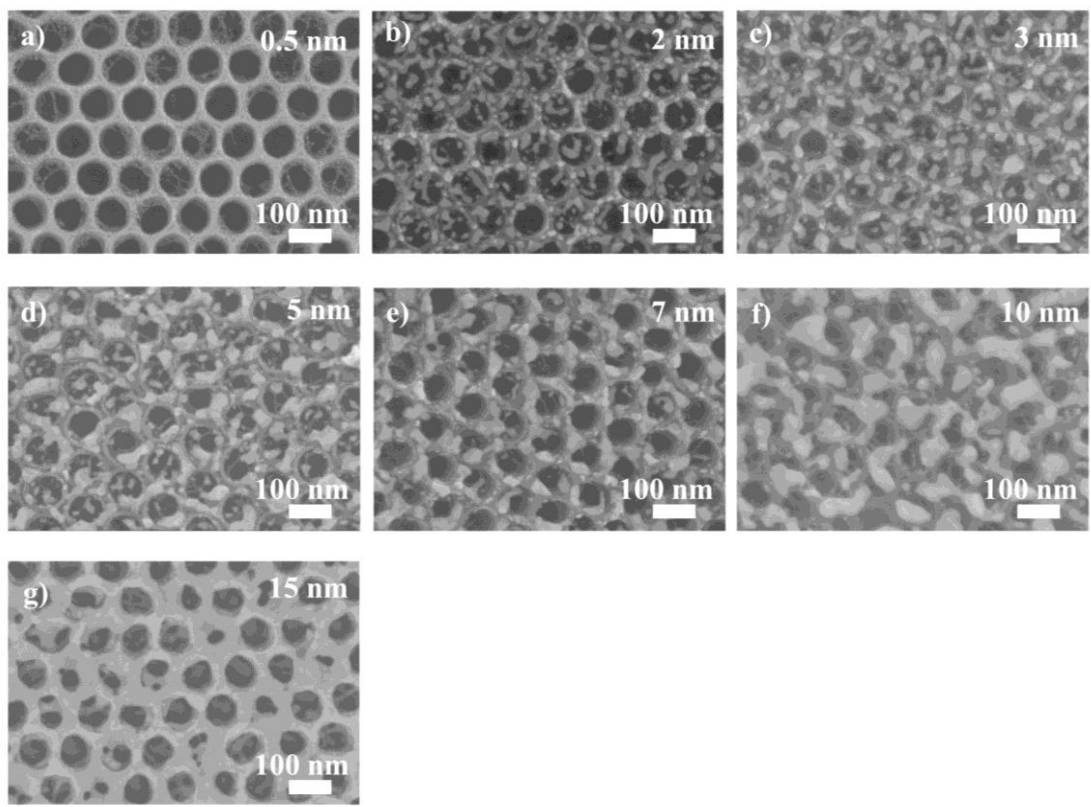

**Figure S3.** Top view SEM images of TiO$_2$ nanotubes decorated with different nominal thicknesses of sputtered Pt films: a) 0.5 nm; b) 2 nm; c) 3 nm; d) 5 nm; e) 7 nm; f) 10 nm and g) 15 nm.

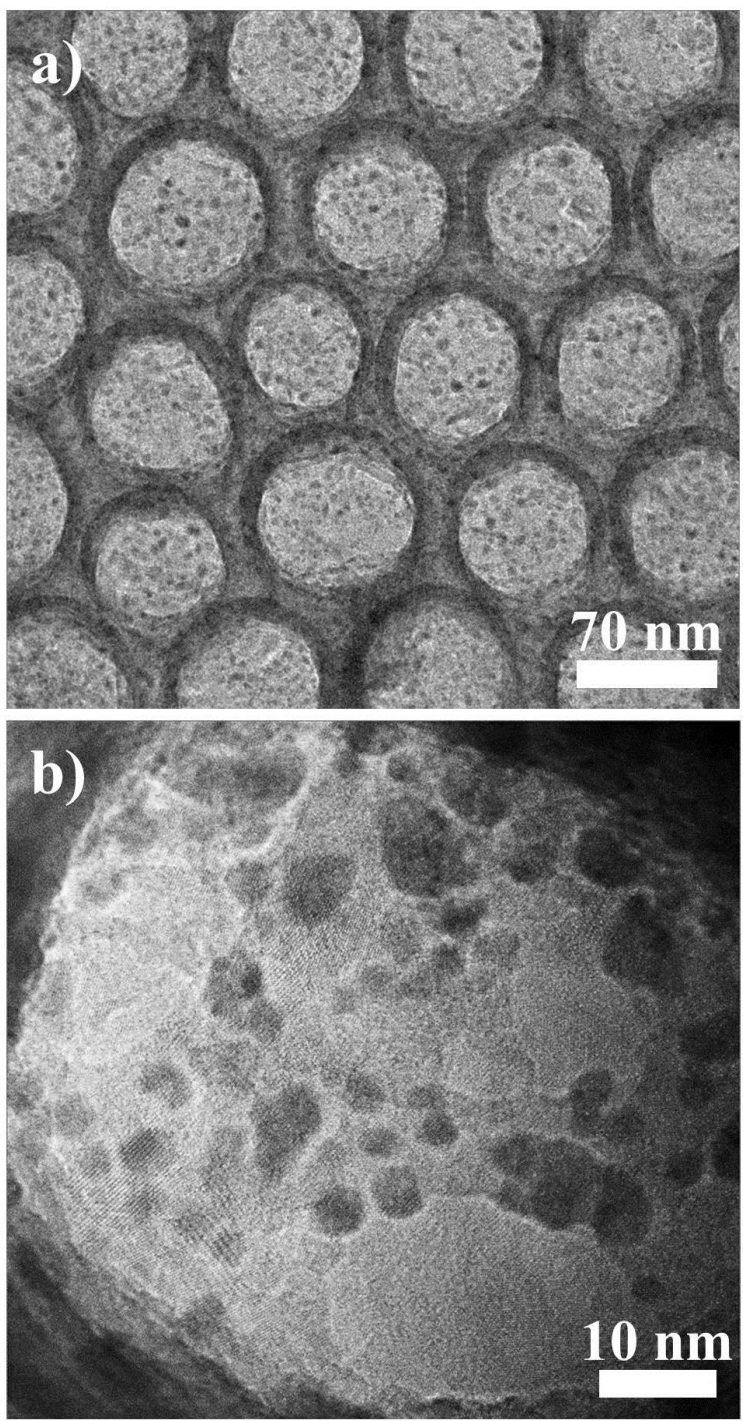

**Figure S4.** HR-TEM images of Pt nanoparticles suspended over TiO$_2$ nanotubes.

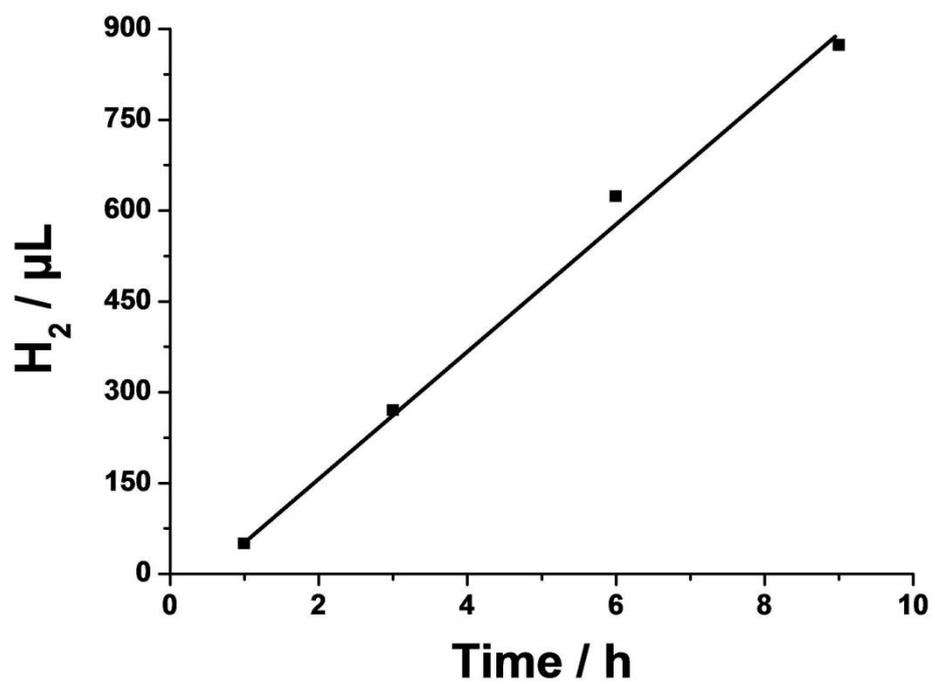

**Figure S5.** Amount of photo-produced $H_2$ measured over the irradiation time with 1 nm-thick Pt suspended over $TiO_2$ nanotubes.

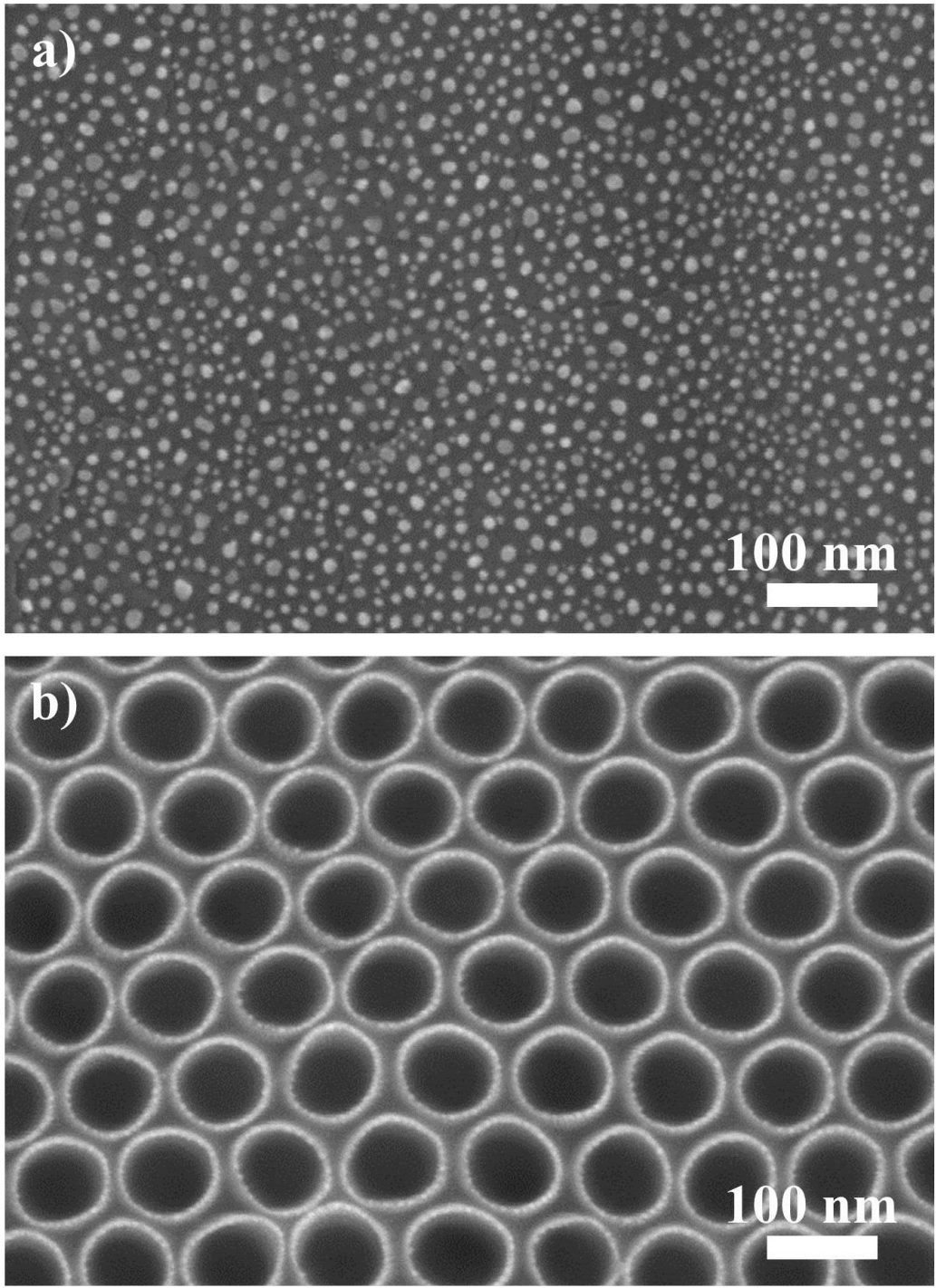

**Figure S6.** Top view SEM images of 1 nm-thick Pt-decorated a) $TiO_2$ compact oxide film (formed by a 15 min-long anodization in 1 M aqueous $H_2SO_4$ at 20 V) and b) $TiO_2$ nanotubes prepared as reported in the experimental but without growing the nanofibrils.

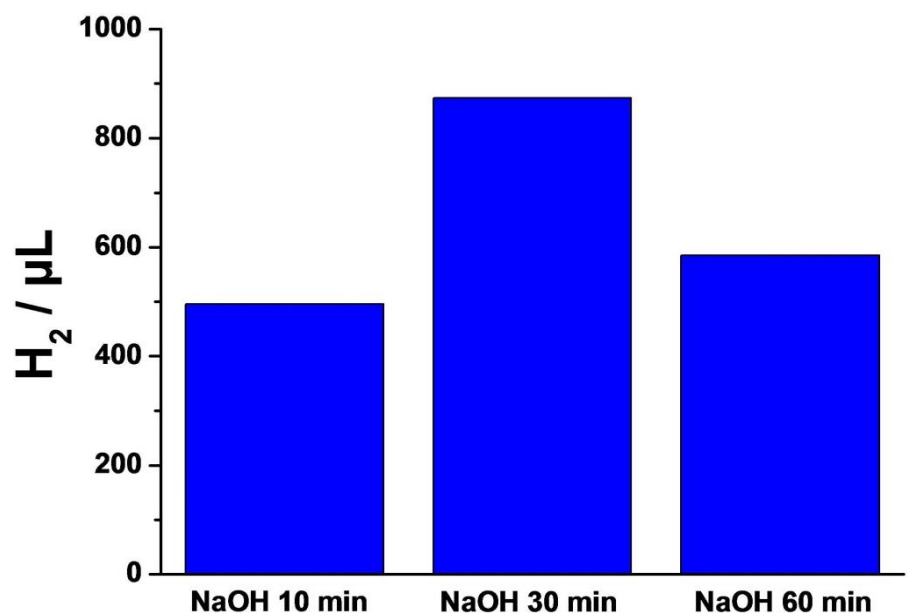

**Figure S7.** Photocatalytic $H_2$ evolution measured with 1 nm-thick Pt suspended over $TiO_2$ nanotubes prepared by different durations of the NaOH-soaking treatment.

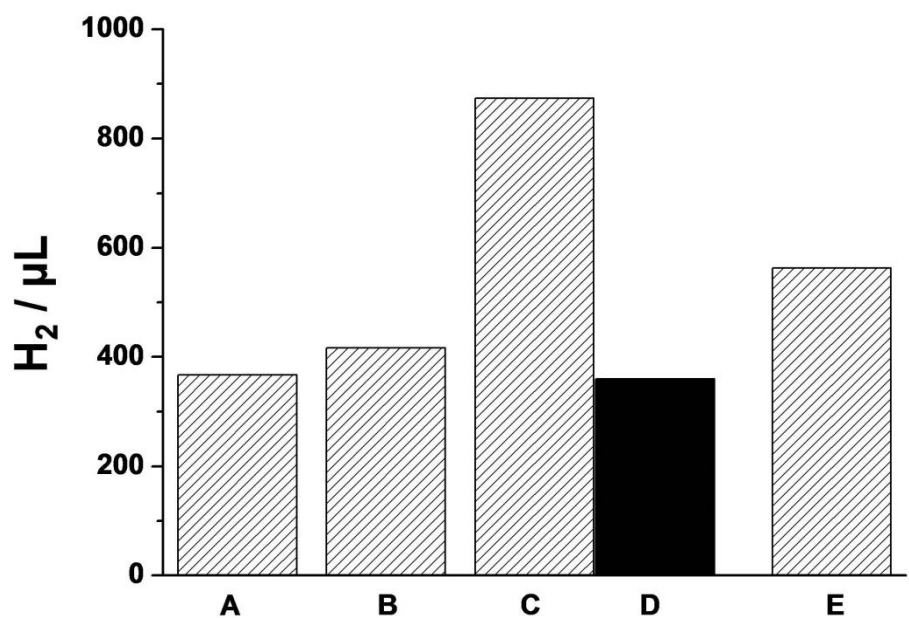

**Figure S8.** Photocatalytic $H_2$ evolution of different 1 nm-thick Pt suspended over $TiO_2$ nanotubes. Samples A, B, C and E were prepared as follows: nanotube growth – air-annealing – NaOH treatment – Pt sputtering – thermal dewetting in Ar for 2, 10, 30 and 120 min, respectively. Sample D was prepared as follows: nanotube growth – air-annealing – NaOH treatment – thermal dewetting in Ar – Pt sputtering.